\documentclass[pra,aps,twocolumn,showpacs,showkeys,
 amsmath,amssymb,reprint,floatfix,superscriptaddress]{revtex4-1}

\usepackage{amsmath}
\usepackage{amssymb}
\usepackage{amsfonts}
\usepackage{graphicx}
\usepackage[usenames,dvipsnames]{xcolor}
\usepackage{bm}
\usepackage{hyperref}
\usepackage{physics} % \ket, \braket
\usepackage{comment}
\usepackage{afterpage}
\usepackage[abs]{overpic}

\newcommand{\kVH}{\ensuremath{\bm{k}_{\text{VH}}}}
\newcommand{\nVH}{\ensuremath{n_{\text{VH}}}}
\newcommand{\EVH}{\ensuremath{E_{\text{VH}}}}
\newcommand{\PsiVH}{\ensuremath{\Psi_{\text{VH}}}}
\newcommand{\nBIC}{\ensuremath{n_{\text{BIC}}}}

\newcommand{\Nsites}{\ensuremath{N_{\mathrm{sites}}}}

\begin{document} % Phys.Rev.

% ===========================================================
% ===========================================================

\title{Van Hove bound states in the continuum: \\ Localised subradiant states in finite open lattices}

\author{Jordi Mur-Petit}
\email{jordi.murpetit@physics.ox.ac.uk}
\affiliation{Clarendon Laboratory, University of Oxford, Parks Road,
	Oxford OX1 3PU, United Kingdom}

\author{Rafael A. Molina}
\email{rafael.molina@csic.es}
\affiliation{Instituto de Estructura de la Materia, IEM-CSIC, Serrano 123, Madrid 28006, Spain}

\begin{abstract}
We show that finite lattices with arbitrary boundaries may support large degenerate subspaces, stemming from the underlying translational symmetry of the lattice. 
When the lattice is coupled to an environment, a potentially large number of these states remains weakly or perfectly uncoupled from the environment, realising a new kind of bound states in the continuum. 
These states are strongly localized along particular directions of the lattice which, in the limit of strong coupling to the environment, leads to spatially-localized subradiant states.
\end{abstract}

%\pacs{%
% 73.63.-b % Electronic transport in nanoscale materials and structures
% 05.60.Gg % Quantum transport
% 42.70.Qs % Photonic bandgap materials
% 37.10.Gh % Atom traps and guides 
%}

\maketitle

% ==========================================================
% ==========================================================

\setlength{\belowdisplayskip}{3pt} \setlength{\belowdisplayshortskip}{3pt}
\setlength{\abovedisplayskip}{3pt} \setlength{\abovedisplayshortskip}{3pt}
\widowpenalty10000
\clubpenalty10000

\section{Introduction}\label{sec:intro}

Controlling the spatial and temporal evolution of light, acoustic or matter waves is at the core of a wealth of modern technologies and research. Just as interference between electronic Bloch waves underlies band structure theory in condensed matter, in photonics the interplay between constructive and destructive interference and spatial localisation is the basis of disparate applications such as lasing~\cite{Mekis1999,Kodigala17}, slow light~\cite{Baba2008}, light focusing~\cite{Yu2014}, or frequency up-conversion~\cite{Koshelev2018}, to name a few.
Recent works have also proposed to identify and harness lattice  states strongly coupled to their environment based on lattice symmetries for light storing~\cite{Facchinetti2016},
and as structured environments with anisotropic properties for quantum optics~\cite{Jenkins2017, Galve2018pra, GonzalezTudela2019}

A concept that unifies the ideas of spatial localization and interference is bound states in the continuum (BICs). BICs are solutions of wave equations that are spatially bounded even though their energy lies within the continuum of a system~\cite{Hsu16}.
Their spatially compact nature stems from interference effects on the potential energy landscape through which the wave propagates. Because of this interferometric origin, BICs are closely related to dark states and sub- and super-radiance phenomena~\cite{Lovera2013}.
The original discovery of BICs relied on a fine tuning of the potential landscape~\cite{vonNeumannWigner29}, however 
more recently Refs.~\cite{Bulgakov2006, Moiseyev2009} showed that BICs can be made robust through symmetry arguments.

Exact BICs have a vanishing width and can hence can be regarded as resonances with an infinite quality factor, $Q$. 
An example of this is chiral BICs~\cite{Mur14}, which are BICs supported by the chiral or sublattice symmetry of a finite lattice. Due to the chiral symmetry, a large number of degenerate states can be found at zero energy, only some of which do couple to the environment when the system is connected to a finite number of in-/out-coupling ports or leads.~\cite{Mur14}.
When the sublattice symmetry of the lattice is weakly broken, these BICs evolve into narrow (high-$Q$) resonances~\cite{Koshelev2018} which can be harnessed for narrow band filtering~\cite{Foley14}, or to devise new nonlinear metasurfaces for frequency up-conversion~\cite{Koshelev2019}.
Chiral BICs are also closely related to compact localized eigenstates in Lieb lattices, in that the former are born from the macroscopic degeneracy of zero-energy states in bipartite lattices, while the latter emerge from interferences between the many degenerate states in the Lieb lattice's flat band. A striking demonstration of this effect was reported by Vicencio {\em et al.}, who showed how these states propagate without distortion in a two-dimensional photonic lattice~\cite{Vicencio15}; similar results have been investigated in quasi-one dimensional structures~\cite{Maimati17}, and there is currently high interest in combining ideas of chirality and flat bands~\cite{Leykam2018}.

In this paper, we identify a more general mechanism supporting BICs in finite lattices, without specific requirements on lattice symmetry or the existence of flat bands. The key idea is that of \textit{van Hove degeneracies}, which are directly connected with van Hove singularities (VHSs).
A VHS is a divergence in the density of states (DOS), $\rho(E)$, of an infinite lattice~\cite{VanHove53, Ziman72}. 
The relation between the dispersion relation of the system, $E(\bm{k})$, and its DOS, $\rho(E) = \int dS/|\nabla E(\bm{k})|$, where $dS$ is the isofrequency surface element, implies that VHSs are found at the critical points of the Brillouin zone, where $E(\bm{k})$ has an extremum. 
In a \textit{finite} lattice of suitable geometry, the VHS translates into a high-dimensional subspace of degenerate states at the van Hove energy --- a VHD.
Further, we show that, when the lattice is coupled to an environment, the states in the VHD split into three subspaces: nondissipative (``subradiant'') localized states that we name VH BICs, subradiant delocalized states, and highly-dissipative localized states, which occur as high-$Q$ resonances for weak coupling. 
We analyse the impact of these subspaces on the transmission properties of photonic crystals, and discuss their relation to dark states and guided modes.
Finally, we discuss the connection of VH BICs with the super-radiance transition in finite lattices.

% ==========================================================
% ==========================================================
\section{Van Hove BICs in open finite lattices\label{sec:vanhove}}

% ===========================================================
\begin{figure*}[t]
 \begin{center}
 \includegraphics[height=0.21\textwidth]{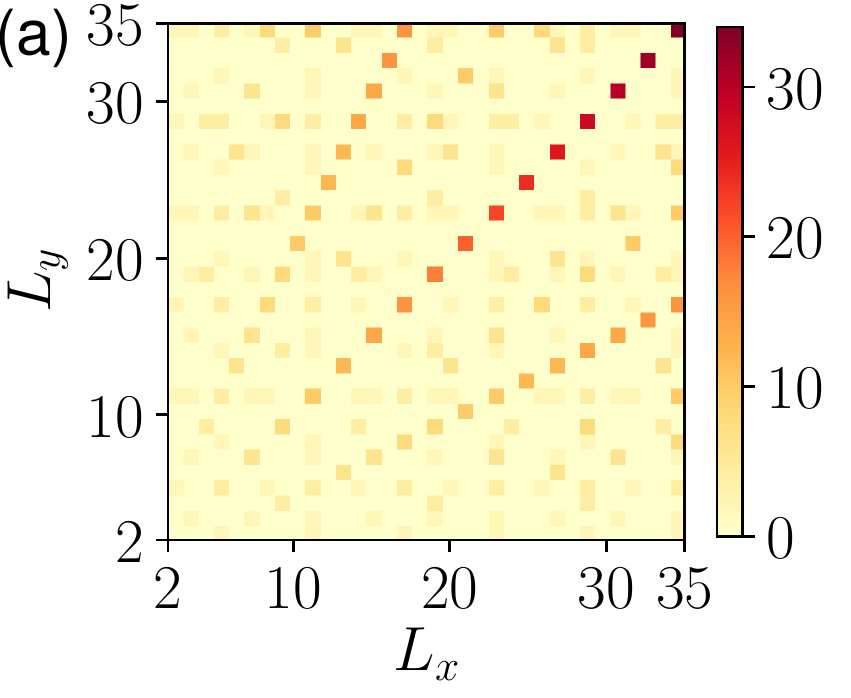}
 \includegraphics[height=0.21\textwidth]{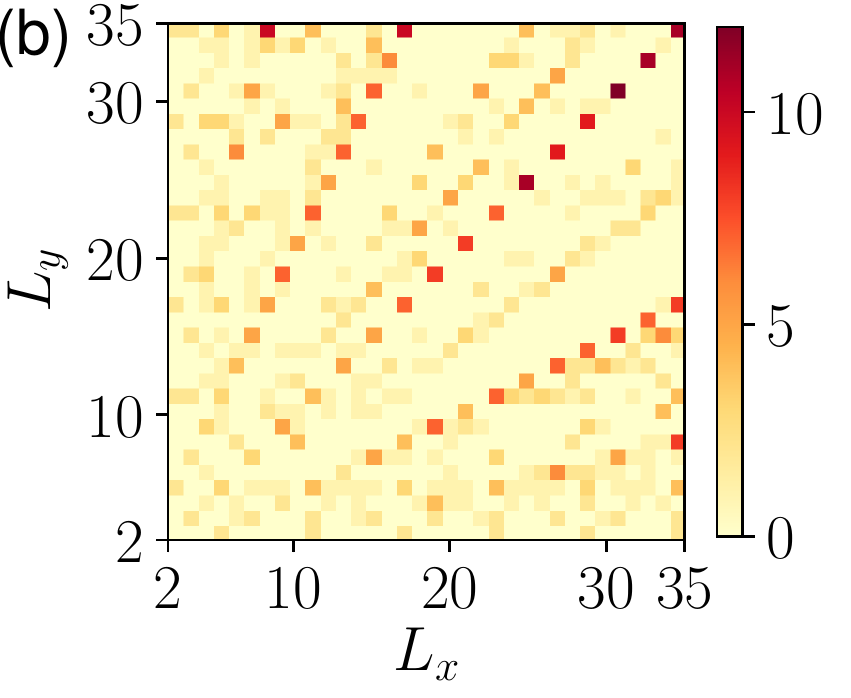} 
 \includegraphics[height=0.21\textwidth]{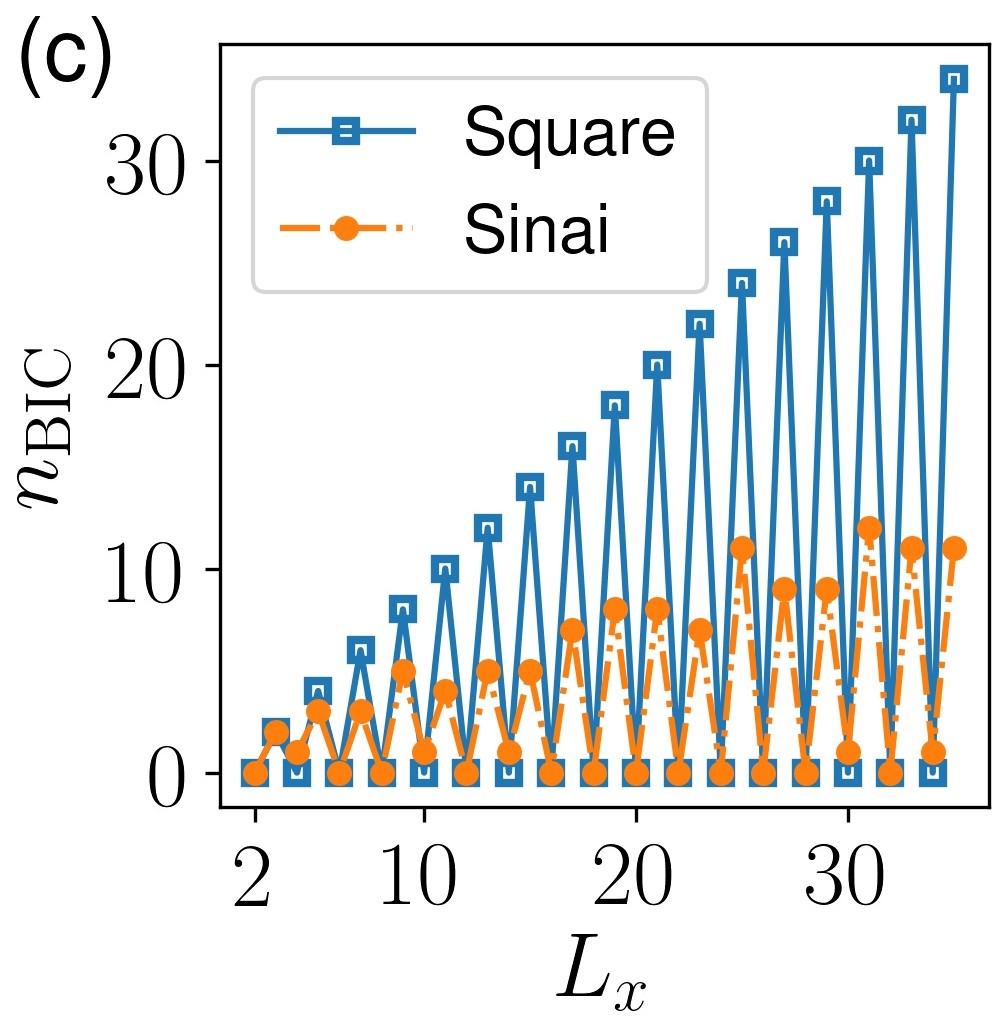} 
 \includegraphics[height=0.21\textwidth]{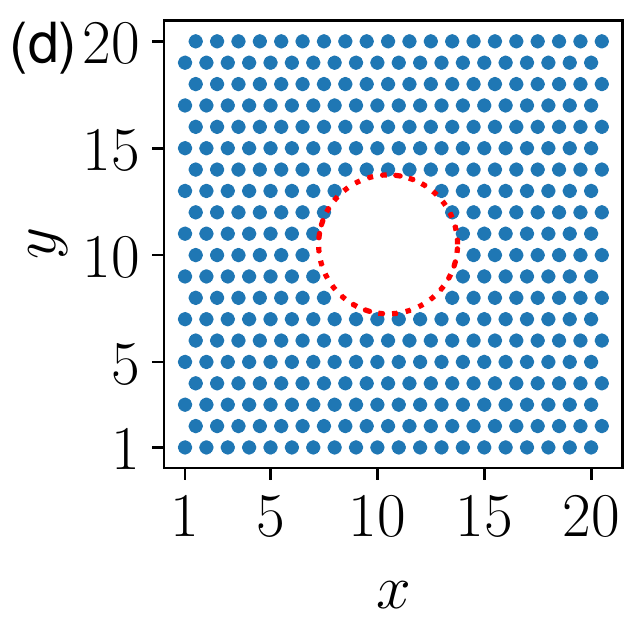} \caption{\label{fig:triangular_degeneracies_sinai}
 (a) 
 Number of degenerate states at the VH energy, $\nVH$, for a triangular lattice with $L_x$ columns and $L_y$ rows. 
 (b) Same as (a) for a triangular lattice with a central hole of radius 1/6 of the smaller size (Sinai billiard). 
 (c) Even-odd effect in $\nVH$ for $L_x\times L_x$ lattices without (blue squares) and with (orange circles) a circular hole at the centre.
 (d) Example $L_x \times L_y = 20 \times 20$ lattice with a circular hole of radius $R=3.25$. Each dot is a lattice site, while the dashed line indicates the boundary of the central hole.}
 \end{center}
\end{figure*}
% ===========================================================

We start considering a tight-binding Hamiltonian $H$ on a two-dimensional lattice, with $L_x$ columns and $L_y$ rows, with hopping between nearest neighbours,
\begin{equation}\label{eq:tight-binding-ham}
 H = -t \sum_{\langle i,j \rangle}
   \left( c_i^{\dagger} c_j + \mathrm{H.c.} \right)  \:,
\end{equation}
where $c_i^\dagger$ creates a particle on site $i$ of the lattice, the sum is over nearest neighbors $\langle i,j\rangle$, and $t$ is the hopping amplitude.
This model can represent a range of material systems including a fiber bundle, where photons hop between single fibers; a set of coupled dielectric resonators placed in a microwave cavity, among which microwave photons hop~\cite{Bellec13}; or single electrons hopping in an array of quantum dots~\cite{Capasso92,Plotnik11}.

For concreteness, we restrict our calculations to a triangular lattice, as this has no sublattice symmetry, thus avoiding complications in the analysis due to the interplay between VH BICs and chiral BICs that appear in the square lattice and the honeycomb~\cite{Mur14}.

In the limit of a large system, $L_x, L_y \gg 1$,
the dispersion relation of this system reads
\begin{equation}
 E(\mathbf{k})
 =-2t \left[ \cos k_x a + 2\cos \frac{k_x a}{2} \cos \frac{\sqrt{3}k_y a}{2} \right] ,
 \label{eq:Ek}
\end{equation}
where $a$ is the lattice constant, $t$ is the nearest-neighbor hoping matrix element, and $\bm{k}=(k_x, k_y)$ is a wave vector in reciprocal space. 
The critical momenta where $|\nabla E(\bm{k})|=0$
fulfill that either $E(\bm{k})=2t$ or the energy is at the upper or lower edges of the spectrum. We are concerned with the former, which correspond to a saddle point of the spectrum and lead to a divergence in the DOS --- a VHS~\cite{VanHove53, Cortes2013}.
These critical momenta are $\kVH^{j}=(k_x,k_y)=(\pm \pi,\pm \pi/\sqrt{3})$, where $j=1,...,4$ indexes the four possible sign combinations.

In a \textit{finite} lattice, only momenta $\mathbf{k}$ compatible with the boundary conditions set by the geometry of the system are possible. 
When the boundaries allow 
stationary waves with wave vectors equal to $\kVH$, the lattice will support a large number of degenerate states at the VH energy, $\EVH = E(\kVH)$, reflecting the singularity that emerges in the infinite-size limit. We refer to this degenerate space associated with the VHS as a VH degenerate subspace or VHD.

The exact dimension, $\nVH$, of the VH degenerate subspace depends in a sensitive manner on the shape of the system.
It can be determined as the number of linearly independent solutions of the Schr\"odinger equation with eigenenergy $\EVH$.
This number equals the dimension of the Hilbert space minus the rank of the square matrix $\mathbb{H} - \EVH \mathbb{I}$, where $\mathbb{H}$ is the Hamiltonian matrix and $\mathbb{I}$ is the $\Nsites\times\Nsites$ identity matrix, with $\Nsites$ the number of lattice sites.
The existence of a degenerate eigenvalue of $\mathbb{H}$ implies that the rank of the Hamiltonian is at most the dimension of $\mathbb{H}$ minus two~\cite{Sutherland1986, Texier02, Mur14, Ramachandran2017}.
Generally, very symmetric structures, such as a square domain ($L_x = L_y$), feature large degeneracies, but arbitrary boundaries may also support high-dimensional VHDs, $\nVH \gg 1$.
The general solution of this problem is highly nontrivial although some results regarding the generalization of Bloch functions to finite lattices are known\cite{Alase2017,Cobanera2018}.

We illustrate this in Fig.~\ref{fig:triangular_degeneracies_sinai}(a), where we show exact-diagonalization results for the dimension of the VHD at $E=2t$ for a triangular lattice with rectangular boundaries, as a function of the length of the sides of the rectangle, $(L_x,L_y)$, with $L_i \in [2, 35]$.
The strong peaks along the diagonals indicate that generally more symmetric structures (when $L_x/L_y$ is a rational number) support larger degeneracies.
We observe a strong even-odd effect [see Fig.~\ref{fig:triangular_degeneracies_sinai}(c)]: square domains with odd-length sides (i.e., $L_x=L_y$ an odd number) systematically support a large number of degenerate states, $\nVH = L_x-1$, while even-length square domains generally support no VH subspaces, $\nVH=0$.

To illustrate the generality of the mechanism supporting the VHD in finite systems, and the fact that it does not require a fine tuning of the lattice geometry, we calculate next $\nVH$ for triangular lattices on rectangular domains with a circular hole in the center of the square [Fig.~\ref{fig:triangular_degeneracies_sinai}(d)]. 
We consider in particular a hole of radius $1/6$ of the smaller dimension, as this corresponds to a discretized version of a Sinai billiard; we designate this system the \textit{Sinai lattice}. This is a well-known chaotic billiard~\cite{Sinai70}, a system that has been studied experimentally with microwave cavities~\cite{Stockmann1990}, and exciton-polaritons in semiconductor microcavities~\cite{Gao2015}, among other photonic systems~\cite{Harayama2011}.
We find that even this chaotic geometry can support large-dimensional VHDs, see Fig.~\ref{fig:triangular_degeneracies_sinai}(b). Occasionally, the VHD of Sinai lattices is even larger than that of full $L_x \times L_y$ lattices; for instance for $L_x=L_y = 10$, the full lattice has $\nVH=0$ (as discussed above), while the Sinai lattice with the same values of $L_x$ and $L_y$ has $\nVH=1$.%
\footnote{Note that for the Sinai lattice, $\nVH$ is not symmetric with respect to the exchange $L_x \leftrightarrow L_y$ because the displacement of neighboring rows with respect to each other in the triangular lattice implies a different number of sites being removed by the hole depending on whether the longest side is in the $x$ or $y$ direction.}

We consider next the fate of the states in the VHD when the system is coupled to an environment. 
When we couple a generic finite lattice to an environment, part of the eigenstates in the VHD will acquire a finite width while part of them will remain spatially bound within the lattice and constitute true BICs, similarly to what happens with chiral BICs~\cite{Victor13,Mur14,Ramachandran2017}.
To prove this, consider the VHD at energy $\EVH$, of dimension $\nVH$.
localized losses at a set of sites $l \in \mathcal{L}$, with loss rate $\gamma_l$, can be described by introducing an effective complex on-site energy, $\gamma_l$, that represents the loss rate at site $l$.
Then, the Schr\"odinger equation for the part of the wave function within the lattice, $\Psi$, becomes
\begin{equation}  
 \mathbb{H}\Psi  -\EVH \Psi -i \sum_{l\in\mathcal{L}} \gamma_l \Psi_l =0.
 \label{eq:Hlossy}
\end{equation}
Here $\Psi_l$ is the component of $\Psi$ at lattice site $l$.
Generally, for each lossy site the rank of the matrix $\mathbb{H}-\EVH\mathbb{I}-\gamma_l \delta_{il}\delta_{jl}$ is increased by one with respect to the rank of the matrix corresponding to the closed system, $\mathbb{H}-\EVH \mathbb{I}$ (there can be exceptions for specific values of $\gamma_l$ where the rank may remain unchanged).
This reduction in the rank corresponds to states in the VHD that acquire a finite width; we refer to them as \textit{VH resonances}. As we will show, these are still strongly localized states, the spatial profiles of which display features directly related to the critical moment $\kVH$, except perhaps in the limit $\gamma \to \infty$.
The other states in the VHD do not acquire a nonzero width but constitute true BICs, which we name \textit{van Hove BICs} (VH BICs).
The number of VH BICs, $\nBIC$, generally satisfies 
$\nBIC = \nVH - |\mathcal{L}|$, where $|\mathcal{L}|$ is the number of lossy sites~\cite{Texier02, Mur14}. 
Given that $\nVH \gg 1$, we have that $\nBIC$ can be very large, depending on the boundary conditions of the finite system.
Finally, the spectrum of the open system contains a third subspace of highly-delocalized states coupled weakly to the environment.

We assess the localization of the eigenstates of Eq.~\eqref{eq:Hlossy} by means of the inverse participation ratio (IPR).
The IPR of a pure state $\Psi$ over a basis set $\{ \ket{i} \}$ is defined as
\begin{align}
   IPR = \frac{1}{\Nsites} \frac{1}{ \sum_i | \braket{i}{\Psi} |^4 } \:.
   \label{eq:IPR}
\end{align}
It quantifies the distribution of $\Psi$ over that set and can be regarded as a measure of the `volume' of the state over the corresponding Hilbert space.
The IPR has been used to analyse localization in various fermionic and bosonic systems, as it is related to the localization length and the (multi)fractal character of a system's eigenstates (see, e.g.,~\cite{Aoki1983, Evers2007, Lindinger2019}.) 
Taking $\{ \ket{i} \}$ as the set of normalised eigenstates with state $\ket{i}$ localized at site $i$ of the lattice, 
a state $\Psi$ that is delocalized throughout the whole lattice, $|\braket{i}{\Psi}|=1/\sqrt{\Nsites}$ $\forall i$, satisfies $IPR_{\text{deloc}} = 1$;
on the other hand, for a state localized on site $j$, i.e., $|\braket{i}{\Psi}|=\delta_{ij}$, one has $IPR_{\text{loc}} = 1/\Nsites$, which tends to zero for large lattices.

% ==========================================================
\begin{figure}[tb]
 \includegraphics[width=\columnwidth]{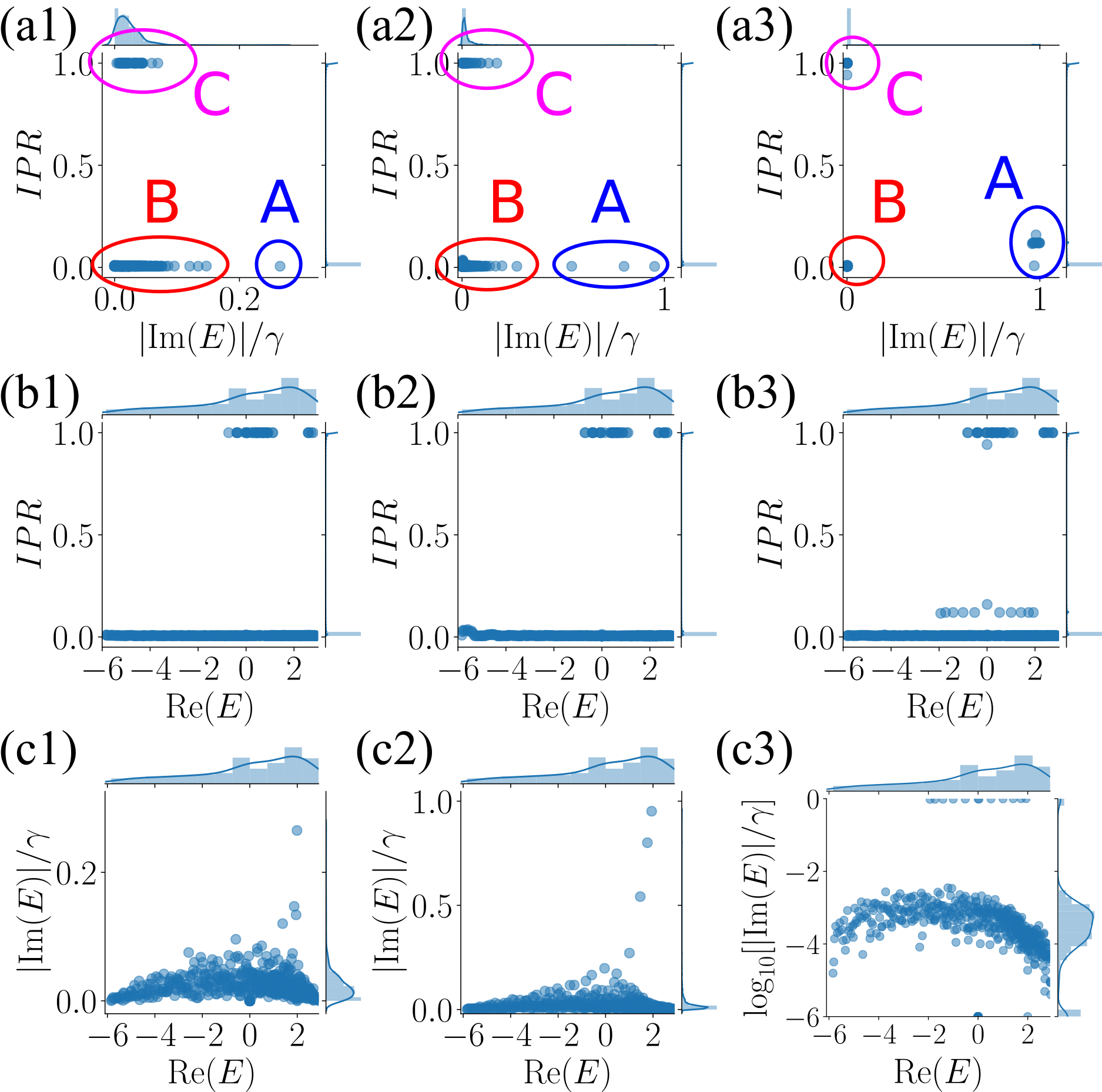}
 \caption{\label{fig:IPRs}
 Properties of the eigenstates of a $23\times23$ Sinai lattice with losses in the 12 left-most sites of the bottom row, with loss rate
 $\gamma=0.102$ (left column), $\gamma=1.002$ (central column), and $\gamma=10.002$ (right column):
 (top row) IPR vs.\ $|\Im(E)|$,
 (middle row) IPR vs.\ $\Re(E)$,
 (bottom row) $|\Im(E)|$ vs.\ $\Re(E)$.
 Labels A, B, C in panels (a1)-(a3) refer to the subspaces defined in Sec.~\ref{sec:vanhove}.
 Curves at the top and right of each panel show the marginal distributions of the data.
 Note panel (c3) is a semilogarithmic plot.
 }
\end{figure}
% ==========================================================

We show in Fig.~\ref{fig:IPRs}(a1) the IPRs of all eigenstates of a $23\times 23$ lattice as a function of the imaginary part of their eigenenergy, for a decay rate $\gamma = 0.102 \ll t $ (we use $t=1$ as our energy unit throughout). Lossy sites with the same value for the coupling strength $\gamma$ are placed in the $11$ leftmost sites of the bottom row. 
We observe that the spectrum is divided into three subspaces:
\begin{itemize}
 \item \textbf{(A)} localized states with $| \Im(E) | \geq \gamma/4$;
 \item \textbf{(B)} localized states with $| \Im(E) | < \gamma/4$;
 \item \textbf{(C)} delocalized states with $| \Im(E) | < \gamma/4$.
\end{itemize}
(Note that the factor $1/4$ on the right hand side stems from observation of the numerical results; we are not aware of an analytical argument for it, e.g., as might be found from a group theory analysis.) 
Figure~\ref{fig:IPRs}(c1) shows that the states with the largest decay rates occur near $\Re(E)=\EVH$. Finally, Fig.~\ref{fig:IPRs}(b1) illustrates the absence of delocalized states ($IPR\approx 1$) near $\EVH$ for small $\gamma$.
The division into the three subspaces A, B, and C as well as their main features remain valid as $\gamma$ is increased. The middle column of Figure~\ref{fig:IPRs} shows the same results for $\gamma \approx t$ while the right column shows results for $\gamma \gg t$.
(For the case $\gamma=0$, the eigenstates of Eq.~\eqref{eq:tight-binding-ham} are generally all delocalised, respecting the translational symmetry and boundary conditions of the lattice.)

Notably, set A of lossy states progressively separates in the $\Im(E)$ axis such that, for $\gamma \gg t$, the few states in set A exhaust the whole set of lossy states and are equal to the number of sites with losses. 
This can be observed in Fig.~\ref{fig:IPRs}(c3), which shows that $\Im(E)$ is very small ($\ll 10^{-2}\gamma$), except for a small number of states, for which $|\Im(E)|/\gamma\simeq 1$.
In Sec.~\ref{sec:superradiance} we will relate this behavior to the phenomenon of superradiance.
Before that, we discuss two types of propagation experiments on finite lattices which enable us to relate the three subspaces to dark states in Sec.~\ref{sec:dark-BICs} and to guided modes in Sec.~\ref{sec:decay_guided_modes}.

% ===========================================
\begin{figure}[tb]
  \includegraphics[width=\columnwidth]{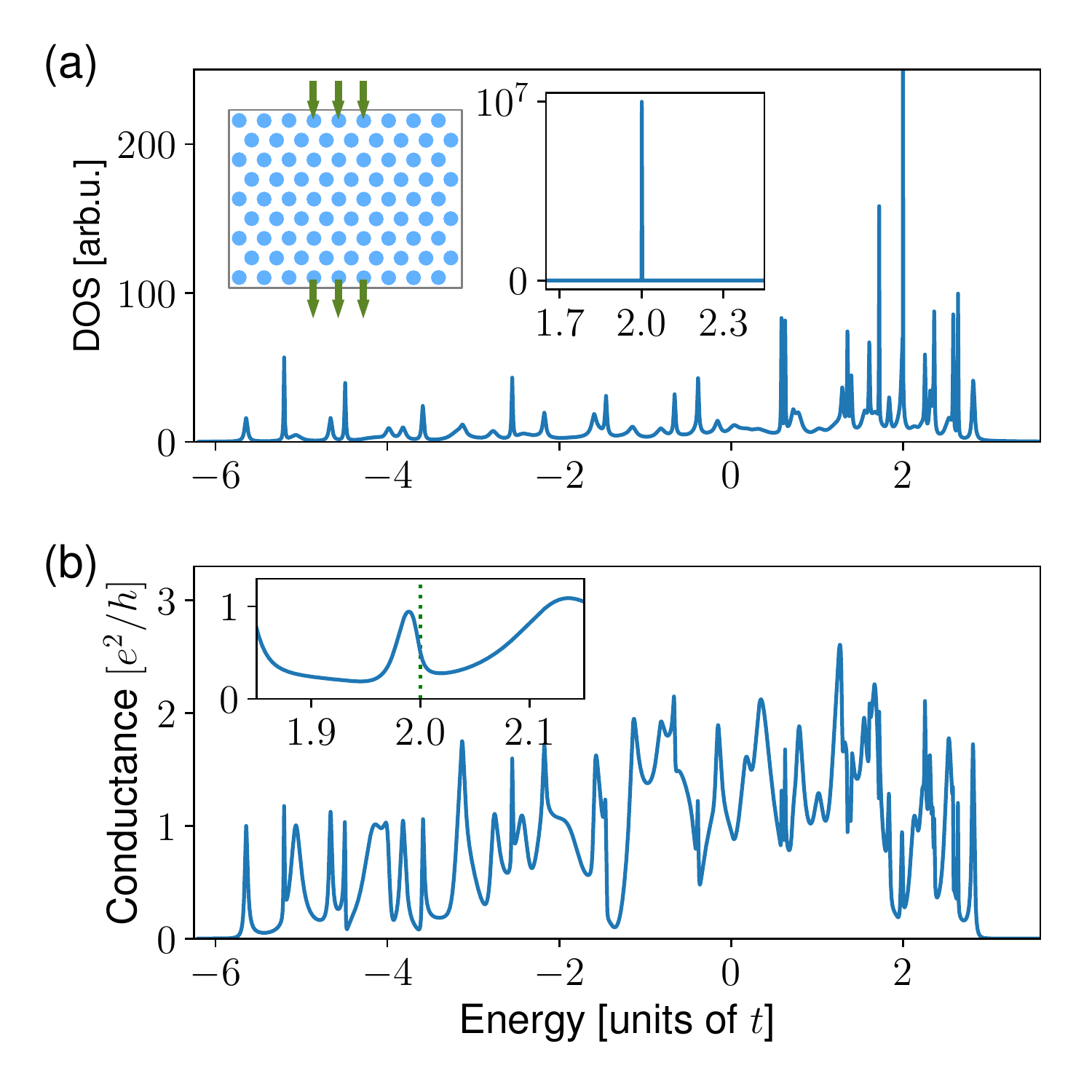}%
 \caption{\label{fig:9x9}
 (a) DOS of a $9 \times 9$ square with connections to fibers on three sites at the top and bottom boundaries as shown in the left inset.
 The right inset shows the DOS divergence close to $\EVH=2t$.
 (b) Transmittance of the same system, normalized to the input flux on a single fiber. The inset highlights the absence of resonant transmission features at $\EVH$.}
\end{figure}
% ===========================================

The qualitative evolution of these three subspaces as we increase the coupling $\gamma$ does not depend on whether $\nBIC$ is finite or zero. 
In the case shown in Fig.~\ref{fig:IPRs}, the number of lossy sites is larger than the dimension of the VHD, but due to the shape of the leaky region, the system supports a small number of lossless states, $\nBIC=7$; however, we have verified that the overall evolution of the spaces A, B, and C with $\gamma$ is the same for a variety of shapes and sizes of the leaky region and sizes of the lattice, which supports the generality of this picture.
Still, we remark that when $\nBIC > 0$, the VH BICs are part of subspace B with $| \Im(E) |$ strictly equal to zero, irrespective of the actual value of the coupling strength $\gamma$.

% ==========================================================
% ==========================================================
\section{Van Hove BICs as dark states in finite lattices}
\label{sec:dark-BICs}

We consider first an experiment connecting leads on two opposite borders of a finite lattice. This corresponds, e.g., to pumping radiation into the lattice through waveguides coupled to selected sites on the top edge of a photonic crystal, and collecting the output radiation through waveguides at the bottom edge [see inset of Fig~\ref{fig:9x9}(a)].

We show in Fig.~\ref{fig:9x9}(a) the DOS for a triangular lattice on a $9 \times 9$ square domain with three-site leads centered on the top and bottom edges; and in Fig.~\ref{fig:9x9}(b) the transmittance of this system calculated using the Landauer-B\"uttiker formalism and Green's functions~\cite{Datta}.

Let us briefly summarise the key ingredients in the calculation using the Landauer-B\"uttiker formalism.
The transmittance, or dimensionless conductance in units of $e^2/h$ for the case of electronic systems, is proportional to the sum of the transmission matrix elements between the left and right leads, $G=T_{RL}(E)$, computed at the energy of interest, e.g., the Fermi energy for electronic conductors at zero temperature. Each of these matrix elements can be computed with matrix elements of the Green's functions of the system with a self-energy coming from the coupling with semi-infinite leads,
\begin{equation}
  G^R(E)=[E\mathbb{I}-H+\Sigma_L+\Sigma_R]^{-1},
\end{equation}
where $\Sigma_L$ and $\Sigma_R$ are the self-energies coming from the left and right leads.
In the absence of inelastic processes, these self-energies can be obtained exactly taking into account the transverse profiles of the modes in each of the leads, $\xi_m$, evaluated at the lead sites coupled to the system sites. Denoting $i$ and $j$ the sites in the system, and $p_i$ and $p_j$ the sites in the lead coupled to them, we have 
\begin{equation}
  \Sigma_{L(R)}(i,j)=-t\sum_m \xi_m(p_i) \exp{+ik_ma}\xi_m (p_j),
\end{equation} 
where $k_m$ is the longitudinal momentum corresponding to the energy of the incoming waves, $a$ is the length scale between sites which we take as $a=1$, and $-t$ is the hopping matrix element between sites in the lead and the system. The transmission matrix element can be expressed in a compact form as:
\begin{equation}
T_{RL}=\mathrm{Tr}[\Gamma_RG^R\Gamma_LG^A],
\end{equation}
where the advanced Green's function is just the Hermitian conjugate of the retarded one, $G^A=[G^R]^{\dagger}$, and the matrix elements of the matrix $\Gamma_p$ ($p=L,R$) for the different modes can be obtained easily from the self-energy and its Hermitian conjugate $\Gamma_p=i(\Sigma_p-\Sigma_p^{\dagger})$ ($p=L,R$). 

As we have three in-/out-coupling channels, the maximum possible transmittance normalized to the flux incoming through one of the input channels is $3$.

Typically the transmittance of such a system is proportional to the DOS, with geometric prefactors depending on the connection to the waveguides.
Thus, a peak in the DOS at an energy $E$ is usually correlated to a peak in the transmission at that energy, although the relative heights of peaks can vary through the spectrum.
This correlation holds generally true in Fig.~\ref{fig:9x9}, except close to $E=\EVH$, where the presence of BICs leads to a divergence of the DOS which, however, does not translate to the transmission properties. 
This indicates that large numbers of states at $\EVH$ do not couple to the leads. This situation is reminiscent of the case of chiral BICs, where the sublattice symmetry leads to an exact zero of the transmittance at zero energy~\cite{Mur14}. However, in the present case the transmittance at $\EVH$ is not exactly zero due to the tails of other peaks close in energy.

When a state in the VHD becomes coupled to the waveguides, its eigenenergy acquires a finite width and the position of the resonance is shifted away from $\EVH$.
For weak coupling, the width acquired is small, and such a state appears as a narrow, or high-$Q$, resonances~\cite{Mur14, Koshelev2018, Koshelev2019}.
Other states in the VHD do not couple to the waveguides and remain 
at $\EVH$; they constitute true BICs (VH BICs), and behave as perfect dark states of the lattice.
Systems with $\nVH \gg 1$ can support a large number of such dark states. The results on $\nVH$ as a function of the lattice geometry [Fig.~\ref{fig:triangular_degeneracies_sinai}] then point to practical guidelines for the design of metasurfaces with a large number of dark states.

% ==========================================================
% ==========================================================
\section{Van Hove BICs as guided modes}
\label{sec:decay_guided_modes}

A complementary propagation experiment consists of injecting a wave packet into the lattice, which is coupled by a number of lossy sites to the environment, and letting it propagate for a long time; this setup models e.g. the propagation of a wave packet down a fiber bundle with some fibers considerably lossier than others.

We model this situation using a non-Hermitian Hamiltonian~\cite{Rotter2009, Moiseyev2011},
\begin{equation}
 \hat{H}_{\mathrm{open}}=\hat{H}-i\gamma \hat{\Gamma},
 \label{eq:Hopen}
\end{equation}
where $\hat{H}$ is the tight-binding Hamiltonian of the closed lattice, and the coupling to the environment is described by the loss rate $\gamma$ and the operator
\begin{equation}
 \hat{\Gamma}=\sum_{j=1}^{N_l} c_j^{\dagger}c_j,
\end{equation}
identifying a number $N_l$ of lossy sites in the lattice.
The invariance of the trace of $\hat{H}_{\mathrm{open}}$ implies a sum rule on the widths of the eigenstates, which here must add to $-N_l\gamma$.

% ======================================================
\begin{figure}[t]
 \includegraphics[width=0.45\columnwidth]{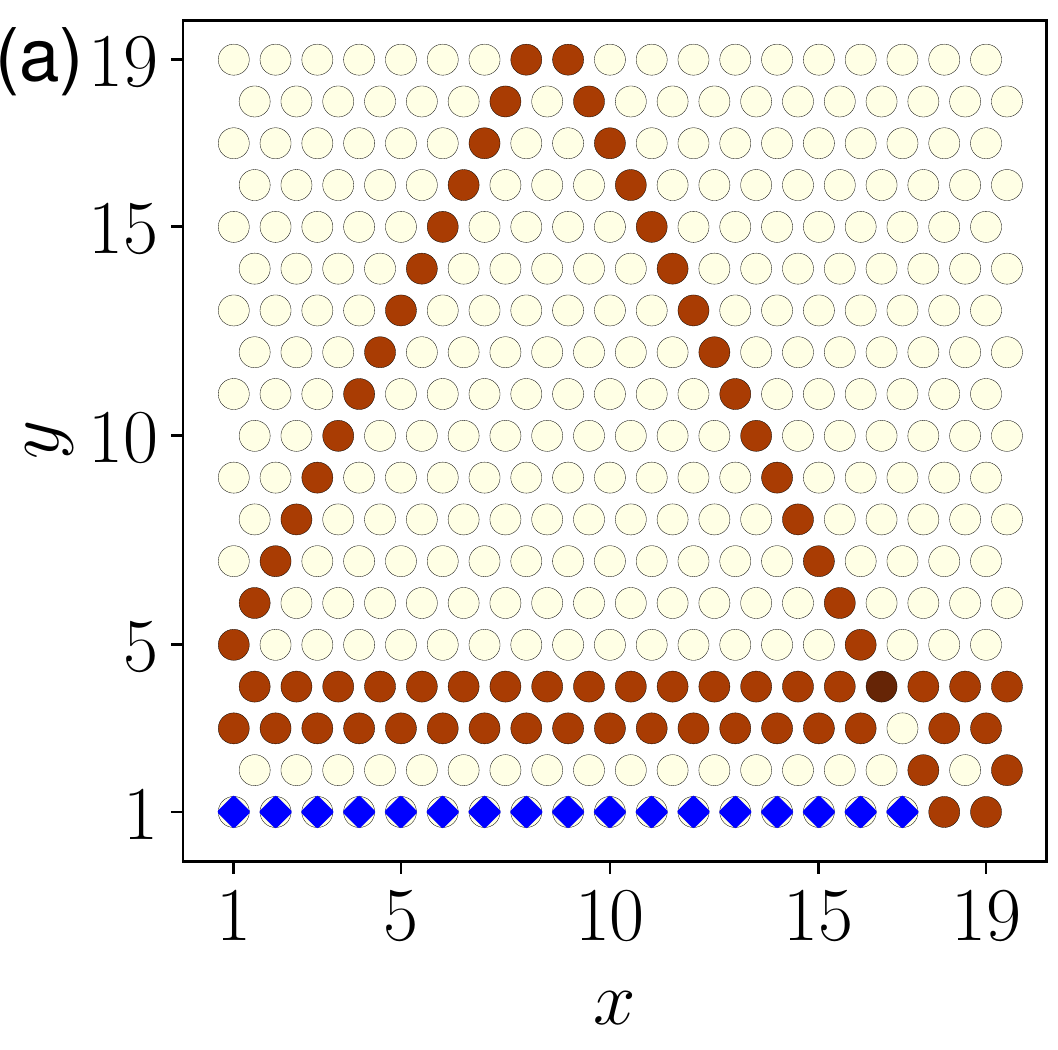}\hfill%
 \includegraphics[width=0.45\columnwidth]{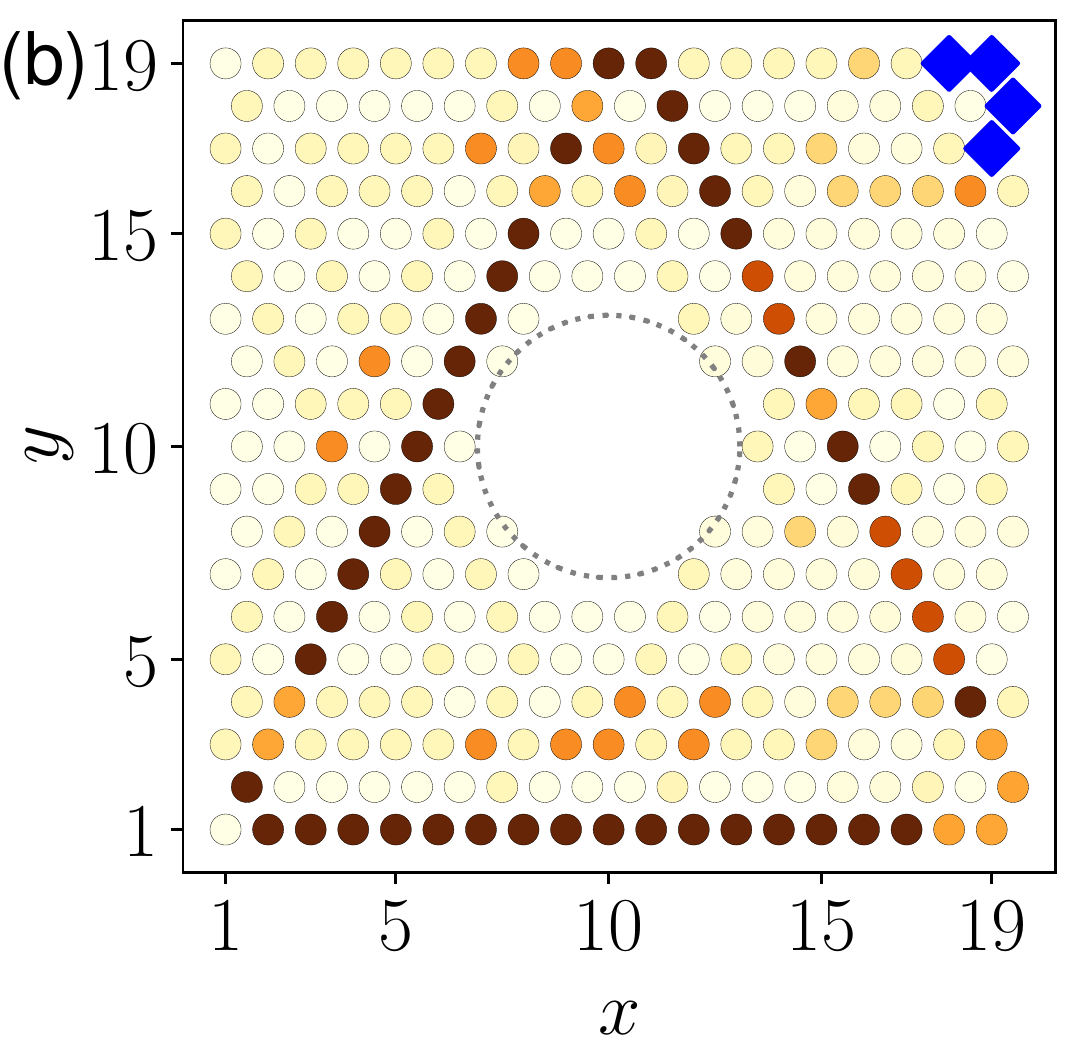}
 \caption{\label{fig:scars}
 Examples of VH lattice scars.
 The intensity of red color is proportional to the probability density at each lattice site (circles) after a long time evolution of a Gaussian wave packet, subject to localized losses at the sites marked by blue diamonds.
 (a) 19 $\times$ 19 square triangular lattice with losses in the $17$ bottom-left sites.  (b) 19 $\times$ 19 Sinai triangular lattice with losses in four sites in the upper right corner.
 The dimensions of the VHD for each closed system are (a) $\nVH=18$ and (b) $5$; only one BIC remains in each system after coupling to the environment.}
\end{figure} 
% =====================================================

We show in Fig.~\ref{fig:scars} the normalised density after a long time evolution simulating one such experiment, with some particular choices of lattice geometry and lossy sites.
Expanding the initial wave packet on the set of eigenstates of Eq.~\eqref{eq:Hopen}, after a long evolution time, the remaining density will be a linear superposition of the eigenstates with smaller $\Im(E)$, the actual density depending on the overlap of the initial wave packet with each state in that space. Given the division of the eigenstates in subspaces A, B, and C, we expect only components within subspaces B and C to survive for long evolution times. Moreover, for a relatively localized initial wave packet, we expect its overlap with highly delocalized states in subspace C to be small. Hence, in practice only the components on the subspace B of VH BICs will be relevant.

The wave function of a state within the VHD can be written
\begin{equation}
 \PsiVH(\bm{r})=\sum_{j} \sum_{\bm{K}} A_{j,\bm{K}}
 e^{i(\kVH^{j}\cdot\bm{r} + \bm{K}\cdot\bm{r})},
 \label{eq:psi}
\end{equation}
where $\bm{r}=(x,y)$ is a position in the lattice, $\kVH^{j}$ are the VH wave vectors of that lattice, and the $\bm{K}$ sum runs over all reciprocal lattice vectors.
For the case of the triangular lattice $\kVH^{j=1..4}=(\pm \pi,\pm \pi/\sqrt{3})$ and $\PsiVH$ will feature increased density, due to constructive interference,
along the diagonals and along zig-zag lines of the lattice, while the destructive interference reduces the density elsewhere. 
This behavior is transmitted to the VH BICs, and hence to the surviving density profile after a long evolution time, as is apparent in Fig.~\ref{fig:scars}.
These localized structures are very much related to the lattice scars observed in chiral lattices~\cite{Victor13} and to the chaotic scars in chaotic quantum billiards~\cite{Heller1984,Sridhar1991} and we name them \textit{VH lattice scars}.
The localization and directionality properties of VH lattice scars can be engineered by designing the spatially varying optical properties of the lattice. 
Conversely, given a lattice geometry, inclusion of lossy sites with controllable decay rate allows triggering a superradiant transition~\cite{Dicke54}, as we discuss next. 

% ==========================================================
% ==========================================================
\section{Van Hove BICs as localized subradiant states}
\label{sec:superradiance}

Several authors have considered the superradiant transition in open systems in the language of non-Hermitian Hamiltonians like Eq.~\eqref{eq:Hopen}, see e.g.\ Ref.~[\citenum{Celardo09, Eleuch2014}].
As the loss rate $\gamma$ is increased, the nature of the system eigenstates evolves as follows. 
For small $\gamma$, the eigenstates of $\hat{H}_{\mathrm{open}}$ are well described by the eigenstates of $\hat{H}$ acquiring some width depending on the amplitude of their wave function at the lossy sites. 
On the other hand, for strong coupling, the eigenstates of $\hat{\Gamma}$ become good eigenstates of $\hat{H}_{\mathrm{open}}$.
In the latter case, for $N_l$ smaller than the number of lattice sites, the set of eigenstates can be divided into `superradiant' states with very large widths $|\Im(E)| \simeq \gamma$ that practically exhaust the sum rule for the resonant width, and `subradiant' states in the kernel of $\hat{\Gamma}$, which remain virtually decoupled from the continuum [$|\Im(E)| \ll \gamma$].
In the language of Sec.~\ref{sec:vanhove}, the `superradiant' states correspond to the highly-dissipative states in subspace A, while the subradiant states are in subspaces B and C.
It follows from the results in Fig.~\ref{fig:IPRs}(a1)-(a3) that all 'superradiant' states are strongly localized ($IPR \ll 1$) for all values of the coupling $\gamma$.
On the other hand, the subradiant states are split between localized and delocalized (extended) states. Note that subradiant states with energy near $\EVH$ are consistently well localized ($IPR \ll 1$) [cf.\ Fig.~\ref{fig:IPRs}(b1)-(b3)].

% ==========================================================
% ==========================================================
\section{Conclusions}\label{sec:conc}

We have established a very general mechanism supporting the existence of bound states in the continuum (BICs) in finite lattices based on van Hove degeneracies, which does not require a distinct symmetry~\cite{Bulgakov2006,Moiseyev2009, Mur14} or flat bands~\cite{Vicencio15, Ramachandran2017}.
At the energy where the density of states diverges due to a van Hove singularity, there may be a large number of degenerate states depending on the boundaries of the finite lattice.
In that case, if the system is coupled to the environment through a finite number of sites, a part of this degenerate subspace generally remains uncoupled from the environment, becoming a set of BICs or localized subradiant states.
These VH BICs are spatially localized along specific direction of the lattice, resembling lattice scars~\cite{Victor13} and compact localized states found in flat bands~\cite{Vicencio15, Maimati17, Leykam2018}.
They do not decay and do not contribute to transport. As such, they could be useful for storing or transmitting information without distortion, e.g., in photonic crystal slabs with losses~\cite{Facchinetti2016,Galve2018}.
Due to the peculiar spatial structure of VH BICs, they could also be harnessed to engineer exotic models relying on a highly-anisotropic coupling between emitters on a photonic crystal~\cite{Wierer2004, Lai2007, Noori2016, Galve2018pra, GonzalezTudela2019}.

On the other hand, states in the VH degenerate space that do couple to the environment are, for weak coupling, spatially localized high-$Q$ resonances that can find applications in slow light and electromagnetically induced transparency (EIT)~\cite{Baba2008,Liu2010}, frequency conversion~\cite{Koshelev2019}, and refractive index sensing (e.g., for biosensing applications)~\cite{Hao2008,Wu2012}, among others~\cite{Khanikaev2013,Limonov2017}. 
For large coupling, these states exhaust all the losses in the system. Because of this, they can be regarded as superradiant states, while the rest of the spectrum, having negligible widths, describes subradiant states, a prediction that could be tested in plasmonic metamaterial arrays at optical frequencies~\cite{Lee2016,Jenkins2017}.

% ==========================================================

\acknowledgments
We acknowledge useful discussions with J. D. Ania-Casta\~{n}\'on and C. Pulido.
This research has been supported by CSIC Research Platform on Quantum Technologies PTI-001, 
UK EPSRC Grant No.\ EP/P01058X/1 (QSUM),
the UK National Quantum Technology Hub in Networked Quantum Information Technologies (Grant No.\ EP/M013243/1),
and the European Research Council under the European Union’s Seventh Framework Programme (FP7/2007-2013)/ERC Grant Agreement No.\ 319286 Q-MAC,
Spain MCIU/AEI/FEDER Project No.\ PGC2018-094180-B-I00,
and CAM/FEDER Project No.\ S2018/TCS-4342 (QUITEMAD-CM).

\bibliography{biblio-bic}

\end{document}